\documentclass[10pt,romanappendices,conference,twocolumn,preprint]{IEEEtran}

\usepackage{amssymb}
\usepackage{graphicx}
\usepackage{amssymb}
\usepackage{amsbsy}
\usepackage{float}
\usepackage{balance}
\usepackage{subfigure}
\usepackage{url}
\usepackage{bm}
\newfloat{longequation}{b}{ext}
\newfloat{longequation}{t}{ext}

\usepackage[cmex10]{amsmath}
\usepackage{array}
\usepackage{varioref}
\usepackage[normalem]{ulem}
\graphicspath{{./figs/}}
\usepackage{cite}
\begin{document}

\title{Multi-User Detection  \\ in  Multibeam Mobile  Satellite Systems: \\ A Fair Performance Evaluation}

\author{\IEEEauthorblockN{ Dimitrios Christopoulos\IEEEauthorrefmark{1}, Symeon Chatzinotas\IEEEauthorrefmark{1}, Jens Krause\IEEEauthorrefmark3  and Bj\"{o}rn Ottersten\IEEEauthorrefmark{1}\IEEEauthorrefmark{2}}
\IEEEauthorblockA{\IEEEauthorrefmark{1}SnT - securityandtrust.lu,  University of Luxembourg
\\email: \textbraceleft dimitrios.christopoulos, symeon.chatzinotas, bjorn.ottersten\textbraceright@uni.lu}
\IEEEauthorblockA{\IEEEauthorrefmark{2}Royal Institute of Technology (KTH), Sweden,
email: bjorn.ottersten@ee.kth.se}
\IEEEauthorblockA{\IEEEauthorrefmark{3}SES, Chateau de Betzdorf, Luxembourg, 
email: jens.krause@ses.com}
}
\maketitle


\begin{abstract}
Multi-User Detection (MUD) techniques are currently being examined as promising  technologies for the next generation of broadband, interactive, multibeam, satellite communication (SatCom) systems. Results in the existing literature have shown that when full frequency and polarization reuse is employed and user signals are jointly processed at the gateway, more than threefold gains in terms of spectral efficiency over conventional systems can be obtained. However, the information theoretic results for the capacity of the multibeam satellite channel, are given under ideal assumptions, disregarding the implementation constraints of such an approach. Considering a real system implementation, the adoption of full resource reuse is bound to increase the payload complexity and power consumption. Since the novel techniques require extra payload resources,  fairness issues in the comparison among the two approaches arise. The present contribution evaluates in a fair manner, the performance of the return link (RL) of a SatCom system serving mobile users that are jointly decoded at the receiver. More specifically, the achievable spectral efficiency of the assumed system is compared to a conventional system under the constraint of equal physical layer resource utilization.
Furthermore, realistic link budgets for the RL of  mobile SatComs are presented, thus allowing the comparison of the systems in terms of achievable throughput. Since the proposed systems operate under the same payload requirements as the conventional systems, the comparison can be regarded as fair.     Finally, existing analytical formulas are also employed to provide closed form descriptions of the performance of clustered multibeam MUD, thus introducing insights on how the performance scales with respect to the  system parameters.

\end{abstract}

\section{Introduction}
  Broadband next generation satellite communication (SatCom) networks are expected to deliver high throughput, interactive services to small and energy efficient, mobile terminals. Novel interference mitigation techniques that have been recently applied to terrestrial networks \cite{3GPP_CoMP}, have proven  a promising  new tool for the design of future SatComs. In satellite networks, information is transmitted either from the broadcaster to the users, i.e. the forward link (FL), or from the users back to the provider, i.e. the return link (RL). Each of these paths is comprised of     two wireless links, namely the uplink, connecting a ground segment (gateway or user terminal) with the satellite and the downlink, vice versa. Focusing on the RL of multibeam satellite networks, Multi-User Detection (MUD) promises substantial gains in terms of spectral efficiency \cite{Christopoulos2011,Christopoulos2012}.
 The key advantage of applying these techniques in SatComs lies in the inherent nature of multibeam satellite networks: a large number of antennas are illuminating a vast coverage area, while all inbound signals are processed in one or more central locations, namely the gateways (GWs). When cooperation amongst the receiving GWs is assumed, MUD over the whole system can be performed\footnote{by the term cooperation, full exchange of channel state information (CSI) and data for all users between the multiple GWs is assumed. When only CSI is exchanged, then the term coordination is commonly adopted.}, over the multiple input multiple output (MIMO) multi-user (MU) multiple access channel (MAC). 

Although promising, the results of the performance of MUD techniques over the multibeam satellite channel are theoretical. Therefore,  the substantial gap between information theoretic results and practical implementation of such approaches needs to be bridged. Two fundamental limitations for the application of such techniques in a real satellite system are the limited capacity of the RL downlink (i.e. the feeder link) and the added on board complexity. Since the feeder link is a point to point link, frequency multiplexing is necessary. Subsequently, an increase in the  user sum-rate, requires proportional increase in the bandwidth of feeder link\footnote{Higher frequency bands (e.g. Q/V bands) for this case are being considered in the design of the future terabit satellite\cite{Mignolo2011}.}. With respect to the on board complexity, when advancing from a specific resource reuse scheme (e.g. 4 color scheme with 2 spectrum segments and 2 polarizations) to full resource reuse, the on board high-power complexity needs to  be  proportionally increased.  
 Current satellites cannot accommodate this payload and the need to explore less complex approaches emanates. Added to that,  fairness issues arise in the comparison of multibeam MUD  to conventional single beam decoding since the first, handles increased physical layer resources compared to the latter.

In the present contribution, a sub-optimal in terms of performance, system design that utilizes exactly the same on board resources, is assumed. The proposed solution is based on  clustering all the co-channel\footnote{by the term co-channel we will refer to beams sharing the shame orthogonal dimension, i.e. identical spectrum segment and polarization. } beams and treating each cluster as a smaller full resource reuse system in which multibeam MUD is employed. Since each cluster is served by one GW in the case of multi-GW systems, the necessity of inter-GW cooperation is also alleviated.
Subsequently, the proposed system
 utilizes the same resources, in terms of payload, feeder link bandwidth and GW interconnection, as conventional systems.

The rest of the present paper is structured as follows. A brief review of the existing related work is provided in Section \ref{sec: Related work}. The considered channel model is described in Section \ref{sec: Channel model}. Section \ref{sec: Capacity analysis}, describes the formulas that provide the spectral efficiency of the considered systems. Finally, the achievable throughput of the proposed system model is calculated through simulations and analytical formulas and compared to the performance of conventional systems, in Section \ref{sec: performance results}. Conclusions are drawn in Section \ref{sec: conclusions}.

\section{Related Work}\label{sec: Related work}

Starting from \cite{Telatar1999},
the capacity of multiple-input multiple-output (MIMO)
systems has been extensively investigated for various types of channels and fading environments, mainly for terrestrial systems. (e.g.\cite{Chatzinotas_JCOM} and the references therein).  Moreover, to cope with the complexity of a hyper-receiver and the difficulty of interconnecting a large number of base stations, the concept of clustered multicell processing has been studied in terrestrial systems \cite{Katranaras2009,Chatzinotas_WCNC}. Similar approaches in the existing literature might also be  referred to as networked MIMO \cite{Zhang2009} or distributed antenna systems (DAS) \cite{Choi2007,Huang2011}.

 Albeit the existing work in terrestrial systems, little is known about the performance of multibeam MUD techniques in SatComs. In this direction, the work of \cite{Letzepis2008} is noted, which elaborated on the uplink capacity of a multibeam satellite system.  The ergodic capacity of correlated MIMO satellite channels was also investigated by \cite{jin2007ergodic}. However, the   composite Rician/lognormal mobile multibeam satellite channel with correlated antennas was proposed in \cite{Christopoulos2011}. Furthermore,  the authors in \cite{Alfano2010} used tools of random matrix theory to upper bound the ergodic capacity and compute the outage probability
of a MIMO Land Mobile Satellite system (LMS). Also, a multiuser decoding algorithm was presented in \cite{Moher2000}, whereas the user mobility has been considered only in \cite{Alfano2010}. Lastly, the performance of linear and non-linear joint processing techniques for the forward and the return link of multibeam SatComs was examined in \cite{Christopoulos2012} while  the concept of GW cooperation for the forward link of multibeam systems has been introduced by the authors in \cite{Zheng2011c}.

To the best of the authors' knowledge, no published results exist for the performance of the return link of clustered multibeam joint decoding systems. Therefore, the novelty of the present work lies in the evaluation of the potential capacity gains of such systems. On this basis, an existing closed-form lower bound for the ergodic capacity of the considered system is additionally employed and extended to provide insight in the impact of the model parameters on the overall performance. In this light, the return link of a mobile multibeam SatCom system  that employs multibeam MUD is hereafter considered;  The underlying fading model is inherently general since it is a combination of small-scale Rician and large-scale lognormal fading.  We note that the latter manifestation is typically caused due to user mobility.  Under these conditions, the performance of the proposed schemes is compared to conventional systems, under equal physical layer resource utilization.

\textit{Notation}: Throughout the paper, \(\mathcal{E}[\cdot]\), \(\left(\cdot\right)^\dag\), denote the expectationand the conjugate transpose of a matrix, respectively.
$\mathbf{I}_n$ denotes an identity matrix of size $n$. Moreover,  \(\mathbf{X}_d=\text{diag}(\mathbf{x})\)  is a diagonal matrix composed of the elements of vector  \(\mathbf{x}=\left[x_1, x_2,\ldots,x_n \right] \).


\section{Composite Fading Multibeam Channel}\label{sec: Channel model}
Let us consider a multibeam satellite scenario. The focus is on the RL uplink  while the feeder link  is assumed perfect but with limited bandwidth. More specifically, a large area covered by \(N\) spot beams is considered, while one single antenna user is scheduled to transmit per beam during a specific time slot. Also, we consider a cluster of \(n\) spot-beams covering \(n\)  user terminals, uniformly distributed, one in each beam. The \(n\) beams are a subset of the total \(N\) beams of the system, selected under criteria to be further explained.  The input-output expression for the \(i\){-th} antenna feed reads as
\begin{equation}\label{eq: General input-output}
y_i= \sqrt{\gamma}\sum_{j=1}^{n} z_{ij}s_j+n_i,
\end{equation}
where \(z_{ij}\) is the complex channel coefficient between the  \(i\)-{th} antenna feed and the  \(j\)-{th} user, $s_j$ is the unit power complex symbol transmitted by each user and  \(n_i\) is the additive white Gaussian noise (AWGN)  measured at
the receive antenna. Since no cooperation among transmit antennas is assumed, every user transmits at the same power $  \gamma$, which
 represents the ratio of the power transmitted by every user over  the equivalent noise power at the receiver. This term is commonly referred to as transmit SNR. To the end of accurately modeling the mobile satellite channel the following characteristics will be incorporated in the channel model: beam gain \(b_{ij}\), log-normal shadowing \(\xi_j\), Rician fading \(h_{j}\) and antenna correlation. Hence, \eqref{eq: General input-output} can be reformulated
\begin{equation}\label{eq: Specific input-output}
y_i= \sqrt{\gamma}\sum_{j=1}^{n} b_{ij}h_{j}\xi_js_j+n_i.
\end{equation}
Fading \(\xi_j\) and $h_j$, only depends on the \(j\)-{th} user position as a result of the practical collocation of the satellite antennas\footnote{Each user sees all the satellite antennas under the same elevation angle. Hence, every user has one corresponding fading instance towards all receive satellite antennas.}. Following from \eqref{eq: General input-output},  the general baseband channel model for all beams  in vectorial form is given by
\begin{equation}
\mathbf{y}= \sqrt{\gamma}\mathbf{Z}\mathbf{x}+\mathbf{n},
\end{equation}
where \(\mathbf{y, x \text{ and } v} \) are \(n\times 1\) receive, transmit, and noise vectors respectively, while, due to \eqref{eq: Specific input-output} the $n \times n$ channel matrix \( \mathbf{Z}\)
will read as
\begin{equation}\label{eq: Z_product}
\mathbf{Z}=\mathbf{B}\mathbf{H}_d\mathbf{\Xi}_d^{1/2},
\end{equation}
where  \(\mathbf{H}_d\) is a diagonal matrix composed  of the Rician fading coefficients and \(\mathbf{\Xi}_d\) also diagonal matrix, with it's entries \(\bm{\xi}=[\xi_1, \xi_2 \dots \xi_n] \) modeled via the classical log-normal distribution. The natural logarithm of these random variables is distributed normaly with  mean $\mu_m$ and variance $\sigma_m$.
Finally, in eq. \eqref{eq: Z_product}, the matrix $\mathbf{B}$ models the satellite antenna gain. It is composed of the square roots of the gain coefficients calculated using the well accepted method of Bessel functions \cite{Diaz2007}.
Further details on he channel model described, are included in \cite{Christopoulos2011}, where the model was initially proposed.

Subsequently, by taking into account the small and large scale fading, a channel model suitable for mobile satellite communications has been presented. 

\subsection{Beam Clustering}
\begin{figure}[ht]
  \centering
  \subfigure[Scenario 1: Beam clustering with conventional frequency and polarization allocation.]{\label{fig: coverage conv}\includegraphics[width=0.19\textwidth]{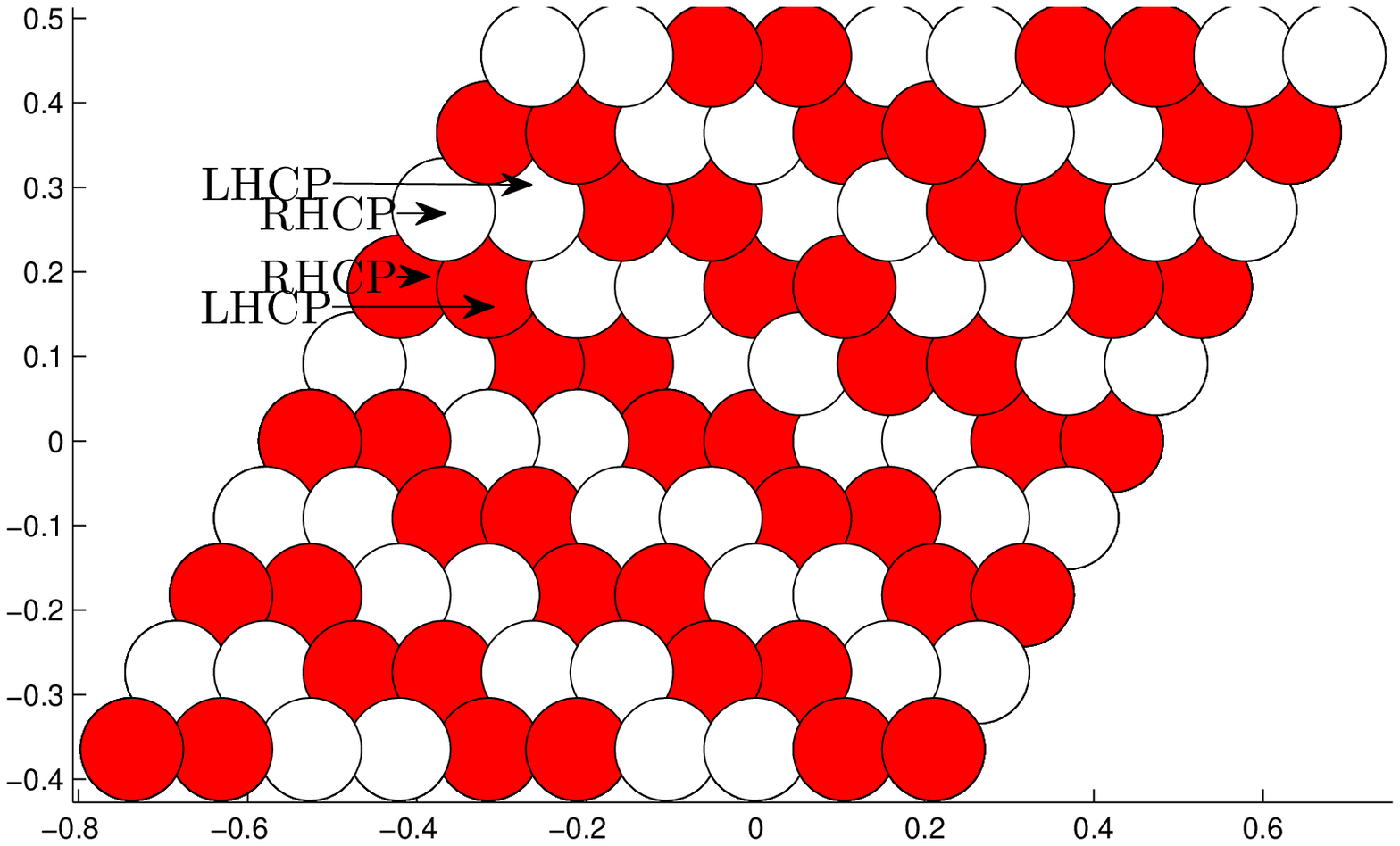}}
  \subfigure[Scenario 2: Beam clustering with adjacent co-channel beams]{\label{fig: coverage clust}\includegraphics[width=0.2\textwidth]{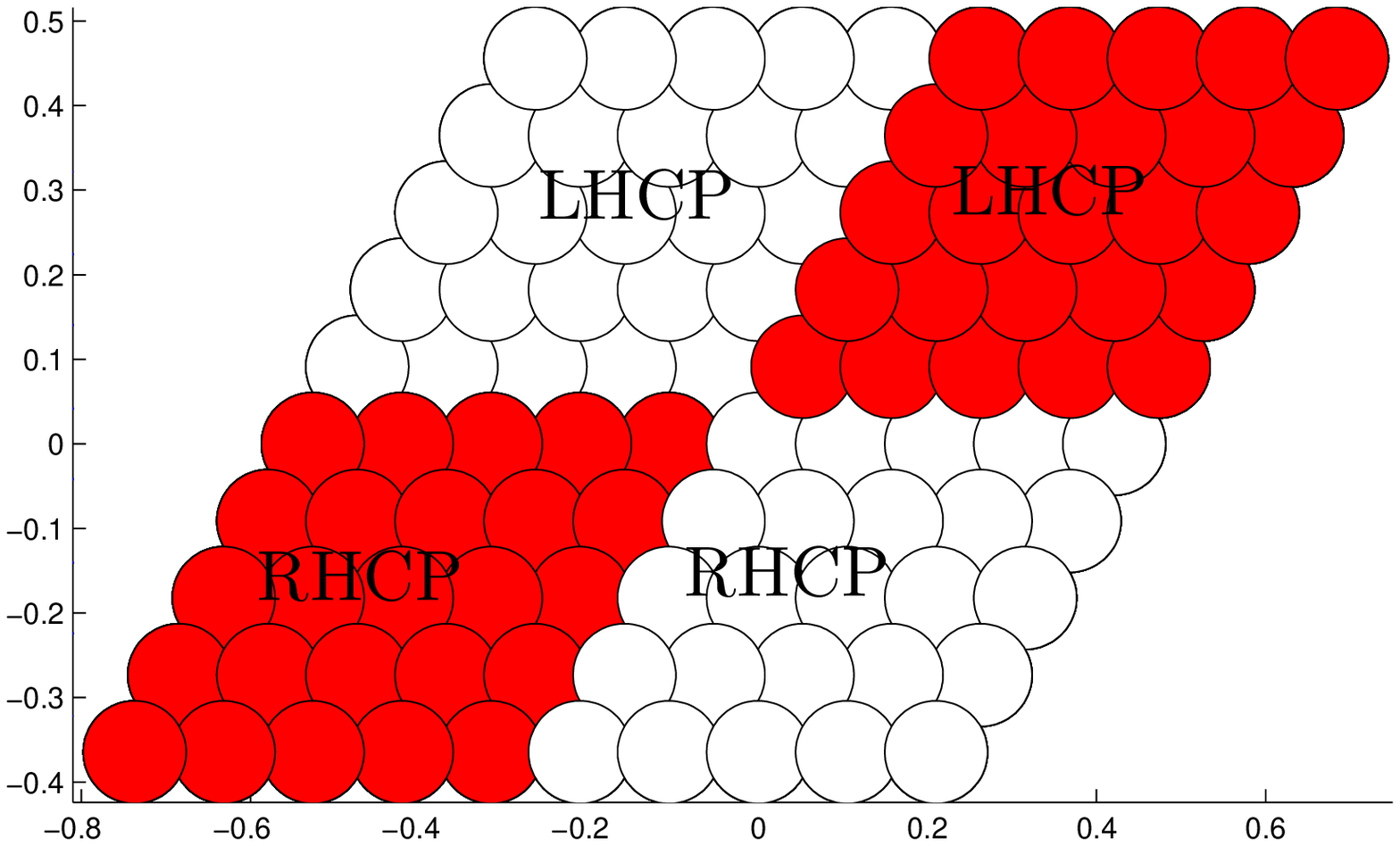}}
  \caption{$N_c=4$ color resource reuse patterns under two different allocation approaches}
  \label{fig:multibeam}
\end{figure}

Beam clustering is achieved by assuming beam gain matrices that contain all the co-channel beams. Consequently, in the first proposed scenario (Fig. \ref{fig: coverage conv}), hereafter referred to as \textit{Scenario 1},  a conventional frequency and polarization allocation pattern is assumed, where the user link bandwidth is divided in two parts  and each one is reused in two polarizations, a Right Hand Circular and a Left Hand Circular Polarization (RHCP and LHCP, respectively). Identical orthogonal dimensions (i.e. colors) are allocated to spatially separated beams, as depicted  in Fig. \ref{fig: coverage conv} and they are handled by the same GW. In the alternative case (\textit{Scenario 2}, depicted in Fig. \ref{fig: coverage clust}) the frequency allocation pattern is altered. All co-channel beams are made adjacent thus increasing the aggregate received power at the receive side. This scenario is inspired from cooperative terrestrial networks where the interferences have a beneficial effect on the system throughput performance\cite{Wyner1994,Somekh2000,Chatzinotas_JCOM}. 
Subsequently, the composite multibeam satellite channel, firstly proposed in \cite{Christopoulos2011} is extended to model the uplink of the clustered multibeam system.
%
\section{Capacity Analysis}\label{sec: Capacity analysis}
\subsection{Optimal Fully Cooperative multibeam system}
In this section, existing results on the capacity of the multibeam composite satellite channel, as achieved by full resource reuse and successive interference cancellation (SIC) at the receive side,  are briefly reviewed. 
To assist the investigation of the multibeam composite Rician/lognormal fading channel, the following analytical lower bound  has  been introduced in \cite{Christopoulos2011}:
\begin{align}
{C}_{\text{erg}}&\geq{N}\log_2\bigg(1+\gamma\exp\bigg(\frac{1}{N} \cdot\ln\left(\det\left(\mathbf{B\cdot B^\dag}\right)\right)+{\mu_m}\notag\\
&-\ln\left(K_r+1\right)+ g_1 \left( K_r\right) \bigg)\bigg),
\label{eq: C_lb_analytical}
\end{align}
 where \(K_r\) is the Rician factor, all other variables as defined in Sec. \ref{sec: Channel model} and  function \(g_1\)  is defined as \cite{Moser2004phd}:
\begin{equation}\label{eq: g_function}
g_{1}\left(s^2\right) \triangleq
 \ln(s^{2})-Ei(-s^{2}),
\end{equation}
where \(Ei(x)\) denotes the exponential integral function.


\subsection{Clustered multibeam system}
In the present contribution, the notion of clustered multibeam joint decoding is analytically modeled. In the proposed system, each cluster of \(n\) beams comprises a SIMO-MU MAC subsystem, operated by a dedicated GW. In this case, the spectral efficiency of the total system is given by
\begin{equation}\label{eq: Log-Det formula clustered}
{C}_{\text{erg}}=\sum_{i=1}^{N_c}\mathcal{E}\left\{\log_2\det\left(\mathrm{I}_{n}+\gamma \mathbf{Z}_i^{\dag}\mathbf{Z}_i\right)\right\},
\end{equation}
where \(\mathbf{Z}_i\) is the channel matrix corresponding to each cluster and \(N_c\) is the number of available orthogonal resources in frequency and polarization (i.e. $N_c=4$ in Fig. \ref{fig:multibeam}). Extending \eqref{eq: C_lb_analytical} for the clustered system case is straightforward: 
\begin{align}
{C}_{ \text{clus}}&=\sum_{i=1}^{N_c}{n}\log_2\bigg(1+\gamma\exp\bigg(\frac{1}{n} \cdot\ln\left(\det\left(\mathbf{B}_i\cdot \mathbf{B}_i^\dag\right)\right)\notag\\
&+{\mu_m}-\ln\left(K_r+1\right)+ g_1 \left(K_r\right) \bigg)\bigg).
\label{eq: C_cluster}
\end{align}

\subsection{Conventional multibeam system }\label{sec: Conventional System}
 The achievable spectral efficiency of conventional multibeam systems employing a frequency reuse scheme is given by
\begin{align}\label{eq: 4c formula}
&{C}_{\text{conv}}=\\
&\mathcal{E}\left\{N_{c}^{-1}\sum_{i=1}^{n}\log_2\left(1+\frac{|z_{ii}|^2}{\sum_{j\neq i,j\in A_C^i}^{}|z_{ij}|^2+ \left(N_{c}\cdot\gamma\right)^{-1}}\right)\right\},\notag
\end{align}
where the channel coefficients \(z_{ij} \) have been defined in \eqref{eq: Specific input-output}, \(\gamma \) is defined in \eqref{eq: General input-output}, \(A^i_C\) is the set of co-channel  to the \(i\)-th, beams and \(N_c\) has been defined in  \eqref{eq: Log-Det formula clustered}.

%

\section{Performance Results}\label{sec: performance results}
\subsection{Simulation Results}
To the end of investigating the potential gains of multibeam joint decoding under realistic and accurate assumptions, parameters from existing S-band Geostationary (GEO) satellite systems are used in the simulation model. 
In Table \ref{tab: fixed link budget}, the link budget of such a system is presented. According to these values, the expressions for the achievable spectral efficiency presented in Section \ref{sec: Capacity analysis}, are employed to calculate the total achievable throughput of the return link.
Monte Carlo simulations (1000  iterations) where carried out to evaluate the average spectral efficiency of the random channels. The results are presented in the following. 

\begin{figure}
               \centering
                    \includegraphics[scale=0.48]{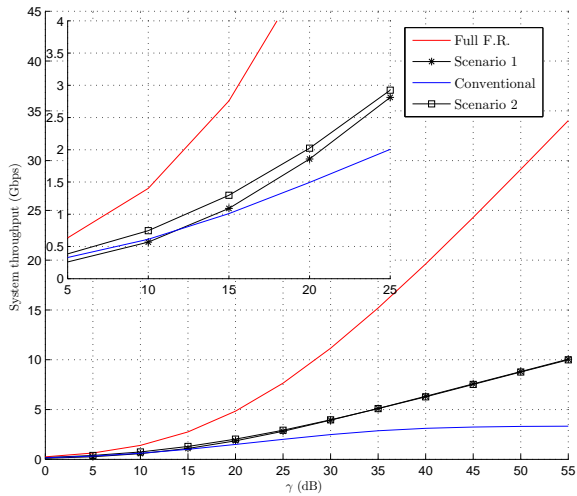}
                    \caption{Performance comparison in terms of achievable throughput between the fully cooperative, the clustered and the conventional systems. }
                    \label{fig: fig2}
                    \end{figure}

According to Fig. \ref{fig: fig2}, the upper bound for the achievable throughput is achieved by  SIC and full resource (i.e. frequency and polarization) reuse. As already proven in \cite{Christopoulos2011,Christopoulos2012}, this capacity achieving decoding strategy can provide more than twofold gain over conventional systems, around the SNR region of operation and even higher gains in as the transmit power increases.  
The performance of the less complex clustered system is presented in the same figure. Results indicate, that in the SNR region of operation of current systems,  i.e. [5--25]dB, a small gain is achieved by the clustered MUD. However, it should be clarified that the comparison is made over equal physical layer resource utilization, while as a technology, it requires minor modifications on existing satellite systems. For larger values of received signal power (i.e. over 30dB), this gain becomes substantial. This originates from the logarithmic dependence of the throughput of conventional systems with respect to $\gamma$ (dB), due to the increase of interferences. Controversially, MUD techniques, exploiting the added spatial degrees of freedom, exhibit linear increase of the throughput with $\gamma$ (dB), in the high SNR region (see Section \ref{sec: analytical approach}), as depicted in Fig.  \ref{fig: fig2}. 

As far as the comparison between the two scenarios of clustered MUD is concerned, the following observations from Fig. \ref{fig: fig2} are made. For less than 25dB, the system proposed in \textit{Scenario 2} marginally outperforms the one proposed in \textit{Scenario 1.  } However, as the received power grows larger, the two schemes perform almost equivalently. Consequently, when comparing the two assumed scenarios, we can conclude that little is gained by reallocating the co-channel beams, in the low transmit power region.   Although intuitively unexpected, since the increase of interferences should significantly increase the performance of joint decoding techniques according to \cite{Wyner1994,Somekh2000,Chatzinotas_JCOM}, this result is in accordance with the more realistic modeling approaches. As proven in \cite{Letzepis2008}, when all interfering tiers are taken into account in the system model, as it is the case in the present work, the increase of interferences leads to a degradation of the system performance. This is especially observed in the high SNR region. Thus, the present  results are explicitly justified. Consequently, if the clustered MUD systems maintain the same levels of transmit power as existing systems, then the beam reallocation should be considered to offer some gain. Otherwise, if power is to be increased, then the beam pattern should follow the conventional standards.


\subsection{Analytical Approach}\label{sec: analytical approach}
\begin{figure}
               \centering
                    \includegraphics[scale=0.48]{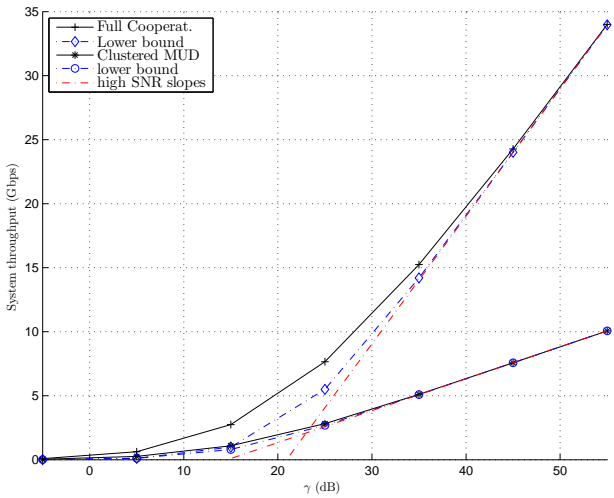}
                    \caption{Analytical lower bounds and high SNR asymptotes  }
                    \label{fig: analytical}
                    \end{figure}

The results presented hitherto, were based on MC simulations.
In order to gain more insight on the implications of the system parameters, analytical calculations firstly introduced in \cite{Christopoulos2011} are extended and employed. Eq. \eqref{eq: C_lb_analytical} is used to lower bound the optimal capacity of the full resource reuse system and \eqref{eq: 4c formula}  that of the clustered system. Results are presented in fig \ref{fig: analytical}. As far as the the full frequency reuse system is concerned, the bound proves exact in the high SNR region but becomes loose in the region [5--25]dB. However, it never exceeds the the theoretical performance, thus it can be employed if worst case results need to be produced. Furthermore, an important insight provided by these formulas is the high SNR slope of each system, evaluating the exact throughput for higher values of $\gamma$. For example, the throughput of the fully cooperative system will behave exactly as
\begin{align}
C_{\text{h-SNR}}& = N\log_2\left(\gamma\right) +\frac{1}{\ln2}\bigg(\ln\big(\det\left(\mathbf{B\cdot B^\dag}\right)\big)+N\mu_m\notag\\
&-N\cdot\big(\ln\left( K_r+1\right)-g_1 \left(K_{r}\right)\big)\bigg).
\label{eq: high SNR slope}
\end{align}
The above high SNR slope shows that the capacity grows linearly with respect to $\gamma$ (in dB) and the inclination of the slope is calibrated by the number of users. Subsequently, the large\ number of the jointly decoded users, benefits the system.  
Also, the negative effect of the high line of sight is also notable by taking into account that $g_1(K_r) =\ln(K_r)-Ei(K_r) $ is positive and increasing with $K_r$. The same observations are made for the clustered MUD system, as presented in Fig. 
\ref{fig: analytical}.

\begin{table}
\caption{ Link Budget }
\centering
\begin{tabular}{l|c}
\textbf{Parameter} & \textbf{MSS}  \\\hline
Orbit &   GEO \\
 Frequency Band   & S (2.2~GHz) \\
 User Link Bandwidth    & 15~MHz  \\
 Number of Beams \(N\) & 100 \\
 Polarizations&RHCP, LHCP\\
 Number of Colors \(N_c\)& 4\\
 Size of Cluster \(n\) &25\\
 Rician factor   \(K\) &13~dB \\
 Lognormal Shadowing $\mu_{m}, \sigma_{m}$ & -2.62, 1.6   \\
Terminal RF power & 4.5 dBw \\
 Receiver noise power    & -133 dBw  \\
 Free Space Loss   & 190~dB \\
 Atmospheric Loss   & 0.5 dB\\
 Tx Antenna Gain    & 3 dB \\
 Max satellite antenna gain  & 52~dBi\\
 Fading Margin    & 3 dB \\
  SNR  region of interest & [-5--25]~dB \\\hline

\end{tabular}
\label{tab: fixed link budget}
\end{table}


\section{Conclusions  }\label{sec: conclusions}
A fair performance comparison between multibeam joint processing and conventional systems, assuming identical physical layer resources, has be elaborated. In order to compare equivalently demanding in terms of resources systems, two clustered MUD schemes are assumed. This distinction is based on the spatial allocation of co-channel beams that form a cluster.  In both cases, co-channel beams are processed by the same GW and MUD techniques are employed to mitigate inter-cluster interferences. However,  interferences originating from beams that are served by a different GW are suppressed by frequency orthogonalization. Consequently, the proposed systems do not employ full resource reuse and demand equivalent to the existing systems, payload complexity, thus making the performance comparison amongst them, fair.
   
The results of the comparison show that clustered MUD can achieve a substantial gain in terms of throughput in the high  SNR region, where the performance of the conventional  schemes saturates due to interferences. It is therefore concluded that in order to maintain the same satellite payload, the receive signal power needs to be increased in order to achieve approximately twofold gain over the conventional systems. In this region, the most preferable approach would be to spatially separate co-channel beams.  

Consequently, the adoption of multibeam MUD techniques in next generation satellite systems can provide substantial gains provided that there is an increase in the transmit signal power. Otherwise, new payloads that can accommodate full resource reuse need to be designed.

%
%

\section*{Acknowledgment}
This work was   partially supported by the National Research Fund, Luxembourg under the project  ``$CO^{2}SAT:$ Cooperative \& Cognitive Architectures for Satellite Networks'.
\bibliographystyle{IEEEtran}
\bibliography{refs/IEEEabrv,refs/conferences,refs/journals,refs/books,refs/references,refs/csi,refs/thesis}

\begin{thebibliography}{10}
\providecommand{\url}[1]{#1}
\csname url@samestyle\endcsname
\providecommand{\newblock}{\relax}
\providecommand{\bibinfo}[2]{#2}
\providecommand{\BIBentrySTDinterwordspacing}{\spaceskip=0pt\relax}
\providecommand{\BIBentryALTinterwordstretchfactor}{4}
\providecommand{\BIBentryALTinterwordspacing}{\spaceskip=\fontdimen2\font plus
\BIBentryALTinterwordstretchfactor\fontdimen3\font minus
  \fontdimen4\font\relax}
\providecommand{\BIBforeignlanguage}[2]{{%
\expandafter\ifx\csname l@#1\endcsname\relax
\typeout{** WARNING: IEEEtran.bst: No hyphenation pattern has been}%
\typeout{** loaded for the language `#1'. Using the pattern for}%
\typeout{** the default language instead.}%
\else
\language=\csname l@#1\endcsname
\fi
#2}}
\providecommand{\BIBdecl}{\relax}
\BIBdecl

\bibitem{3GPP_CoMP}
{3GPP}, ``Further advancements for {E-UTRA} physical layer aspects,'' 3rd
  Generation Partnership Project, {Tech. Rep. TR 36.814}.

\bibitem{Christopoulos2011}
D.~Christopoulos, S.~Chatzinotas, M.~Matthaiou, and B.~Ottersten, ``Capacity
  analysis of multibeam joint decoding over composite satellite channels,'' in
  \emph{Proc. of 45th Asilomar Conf. on Signals, Systems and Computers},
  Pacific Grove, CA, Nov. 2011, pp. 1795 --1799.

\bibitem{Christopoulos2012}
\BIBentryALTinterwordspacing
D.~Christopoulos, S.~Chatzinotas, G.~Zheng, J.~Grotz, and B.~Ottersten,
  ``Linear and non-linear techniques for multibeam joint processing in
  satellite communications,'' \emph{EURASIP J. on Wirel. Commun. and Networking
  2012, 2012:162}. [Online]. Available:
  \url{http://jwcn.eurasipjournals.com/content/2012/1/162}
\BIBentrySTDinterwordspacing

\bibitem{Mignolo2011}
D.~Mignolo, R.~Emiliano, A.~Ginesi, A.~B. Alamanac, P.~Angeletti, and
  M.~Harverson, ``Approaching terabit/s satellite capacity: A system
  analysis,'' in \emph{Proc. Ka Broadband Conf.}, Oct. 2011.

\bibitem{Telatar1999}
I.~E. Telatar, ``Capacity of multi-antenna {Gaussian} channels,'' \emph{Eur.
  Trans. Telecommun.}, vol.~10, no.~6, pp. 585--595, Nov. 1999.

\bibitem{Chatzinotas_JCOM}
S.~Chatzinotas, M.~Imran, and R.~Hoshyar, ``Multicell decoding sum-rate and
  user-rate shares of the cellular {MIMO} uplink channel,'' \emph{{IEEE} Trans.
  Commun.}, 2009, submitted.

\bibitem{Katranaras2009}
E.~Katranaras, M.~Imran, and R.~Hoshyar, ``Sum rate of linear cellular systems
  with clustered joint processing,'' in \emph{Vehicular Technology Conference,
  2009. VTC Spring 2009. IEEE 69th}, 26-29 2009, pp. 1 --5.

\bibitem{Chatzinotas_WCNC}
S.~Chatzinotas and B.~Ottersten, ``Interference alignment for clustered
  multicell joint decoding,'' in \emph{Wireless Communications and Networking
  Conference (WCNC), 2011 IEEE}, Mar. 2011, pp. 1966 --1971.

\bibitem{Zhang2009}
J.~Zhang, R.~Chen, J.~Andrews, A.~Ghosh, and R.~Heath, ``Networked mimo with
  clustered linear precoding,'' \emph{Wireless Communications, IEEE
  Transactions on}, vol.~8, no.~4, pp. 1910 --1921, Apr. 2009.

\bibitem{Choi2007}
W.~Choi and J.~Andrews, ``Downlink performance and capacity of distributed
  antenna systems in a multicell environment,'' \emph{{IEEE} Trans. Wireless
  Commun.}, vol.~6, no.~1, pp. 69--73, Jan 2007.

\bibitem{Huang2011}
Y.~Huang, G.~Zheng, M.~Bengtsson, K.-K. Wong, L.~Yang, and B.~Ottersten,
  ``Distributed multicell beamforming with limited intercell coordination,''
  \emph{{IEEE} Trans. Signal Process.}, vol.~59, no.~2, pp. 728 --738, feb.
  2011.

\bibitem{Letzepis2008}
N.~Letzepis and A.~Grant, ``Capacity of the multiple spot beam satellite
  channel with {R}ician fading,'' \emph{{IEEE} Trans. Inf. Theory}, vol.~54,
  no.~11, pp. 5210--5222, Nov. 2008.

\bibitem{jin2007ergodic}
S.~Jin, X.~Gao, and X.~You, ``{On the ergodic capacity of rank-1 Ricean-fading
  MIMO channels},'' \emph{{IEEE} Trans. Inf. Theory}, vol.~53, no.~2, pp.
  502--517, Feb. 2007.

\bibitem{Alfano2010}
G.~Alfano, A.~De~Maio, and A.~M. Tulino, ``A theoretical framework for {LMS}
  {MIMO} communication systems performance analysis,'' \emph{{IEEE} Trans. Inf.
  Theory}, vol.~56, no.~11, pp. 5614--5630, Nov. 2010.

\bibitem{Moher2000}
M.~Moher, ``Multiuser decoding for multibeam systems,'' \emph{{IEEE} Trans.
  Veh. Technol.}, vol.~49, no.~4, pp. 1226--1234, Jul. 2000.

\bibitem{Zheng2011c}
G.~{Zheng}, S.~{Chatzinotas}, and B.~{Ottersten}, ``Multi-gateway cooperation
  in multibeam satellite systems,'' in \emph{Proc. of 23rd IEEE symp. on
  Person. Indoor Mob. Radio Commun.}, 2012.

\bibitem{Diaz2007}
M.~Diaz, N.~Courville, C.~Mosquera, G.~Liva, and G.~Corazza, ``Non-linear
  interference mitigation for broadband multimedia satellite systems,'' in
  \emph{Proc. Int. Work. Sat. Space Commun. (IWSSC)}, Sept. 2007, pp. 61--65.

\bibitem{Wyner1994}
A.~Wyner, ``Shannon-theoretic approach to a {Gaussian} cellular multiple-access
  channel,'' \emph{{IEEE} Trans. Inf. Theory}, vol.~40, no.~6, pp. 1713--1727,
  Nov 1994.

\bibitem{Somekh2000}
O.~Somekh and S.~Shamai, ``Shannon-theoretic approach to a {Gaussian} cellular
  multiple-access channel with fading,'' \emph{{IEEE} Trans. Inf. Theory},
  vol.~46, no.~4, pp. 1401--1425, July 2000.

\bibitem{Moser2004phd}
S.~Moser, ``{Duality-based bounds on channel capacity},'' Ph.D. dissertation,
  Swiss Federal Institute of Technology (ETH), Switzerland, Oct. 2004.

\end{thebibliography}

\end{document}